*Article*

# Applications of Blockchain Technology beyond Cryptocurrency

**Mahdi H. Miraz[1, *], Maaruf Ali[2]**

[1]School of Computer Studies, AMA International University BAHRAIN (AMAIUB), Bahrain
m.miraz@amaiu.edu.bh
[2]Department of Science and Technology, University of Suffolk, Ipswich, Suffolk, UK
m.ali2@uos.ac.uk
*Correspondence: m.miraz@amaiu.edu.bh

**Abstract:** Blockchain (BC), the technology behind the Bitcoin crypto-currency system, is considered to be both alluring and critical for ensuring enhanced security and (in some implementations, non-traceable) privacy for diverse applications in many other domains - including in the Internet of Things (IoT) eco-system. Intensive research is currently being conducted in both academia and industry applying the Blockchain technology in multifarious applications. Proof-of-Work (PoW), a cryptographic puzzle, plays a vital rôle in ensuring BC security by maintaining a digital ledger of transactions, which is considered to be incorruptible. Furthermore, BC uses a changeable Public Key (PK) to record the users' identity, which provides an extra layer of privacy. Not only in cryptocurrency has the successful adoption of BC been implemented but also in multifaceted non-monetary systems such as in: distributed storage systems, proof-of-location, healthcare, decentralized voting and so forth. Recent research articles and projects/applications were surveyed to assess the implementation of BC for enhanced security, to identify associated challenges and to propose solutions for BC enabled enhanced security systems.

## 1. Introduction

The goal of this research paper is to summarise the literature on implementation of the Blockchain and similar digital ledger techniques in various other domains beyond its application to crypto-currency and to draw appropriate conclusions. Blockchain being relatively a new technology, a representative sample of research is presented, spanning over the last ten years, starting from the early work in this field. Different types of usage of Blockchain and other digital ledger techniques, their challenges, applications, security and privacy issues were investigated. Identifying the most propitious direction for future use of Blockchain beyond crypto-currency is the main focus of the review study.

Blockchain (BC), the technology behind Bitcoin crypto-currency system, is considered to be essential for forming the backbone for ensuring enhanced security and privacy for various applications in many other domains including the Internet of Things (IoT) eco-system. International research is currently being conducted in both academia and industry applying Blockchain in varied domains. The Proof-of-Work (PoW) mathematical challenge ensures BC security by maintaining a digital ledger of transactions that is considered to be unalterable. Furthermore, BC uses a changeable

Public Key (PK) to record the users' identity that provides an extra layer of privacy. The successful adoption of BC has been implemented in diverse non-monetary systems such as in online voting, decentralized messaging, distributed cloud storage systems, proof-of-location, healthcare and so forth. Recent research articles and projects/applications were surveyed to ascertain the implementation of BC for enhanced security and to identify its associated challenges and thence to propose solutions for BC enabled enhanced security systems.

The knowledge domain of the research is in the realm of the digital ledger, specifically, in Blockchain and crypto-currency.

## 2. Technology Fundamentals of Blockchain

This section briefly describes the fundamentals of the technology behind the Blockchain. A Blockchain comprises of two different components, as follows:
1. **Transaction:** A transaction, in a Blockchain, represents the action triggered by the participant.
2. **Block:** A block, in a Blockchain, is a collection of data recording the transaction and other associated details such as the correct sequence, timestamp of creation, etc.

The Blockchain can either be public or private, depending on the scope of its use. A public Blockchain enables all the users with read and write permissions such as in Bitcoin, access to it. However, there are some public Blockchains that limit the access to only either to read or to write. On the contrary, a private Blockchain limits the access to selected trusted participants only, with the aim to keep the users' details concealed. This is particularly pertinent amongst governmental institutions and allied sister concerns or their subsidies thereof.

One of the major benefits of the Blockchain is that it and its implementation technology is public. Each participating entities possesses an updated complete record of the transactions and the associated blocks. Thus the data remains unaltered, as any changes will be publicly verifiable. However, the data in the blocks are encrypted by a private key and hence cannot be interpreted by everyone.

Another major advantage of the Blockchain technology is that it is decentralized. It is decentralized in the sense that:
- There is no single device that stores the data (transactions and associated blocks), rather they are distributed among the participants throughout the network supporting the Blockchain.
- The transactions are not subject to approval of any single authority or have to abide by a set of specific rules, thus involving substantial trust as to reach a consensus.
- The overall security of a Blockchain eco-system is another advantage. The system only allows new blocks to be appended. Since the previous blocks are public and distributed, they cannot be altered or revised.

For a new transaction to be added to the existing chain, it has to be validated by all the participants of the relevant Blockchain eco-system. For such a validation and verification process, the participants must apply a specific algorithm. The relevant Blockchain eco-system defines what is perceived as "valid", which may vary from one eco-system to another. A number of transactions, thus approved by the validation and verification process, are bundled together in a block. The newly prepared block is then communicated to all other participating nodes to be appended to the existing chain of blocks. Each succeeding block comprises a hash, a unique digital fingerprint, of the preceding one.

Figure 1 demonstrates how Blockchain transactions takes place, using a step-by-step example. Bob is going to transfer some money to Alice. Once the monetary transaction is initiated and hence triggered by Bob, it is represented as a "transaction" and broadcast to all the involved parties in the networks. The transaction now has to get "approval" as being indeed "valid" by the Blockchain eco-system. Transaction(s) once approved as valid along with the hash of the succeeding block are then fed into a new "block" and communicated to all the participating nodes to be subsequently appended to the existing chain of blocks in the Blockchain digital ledger.

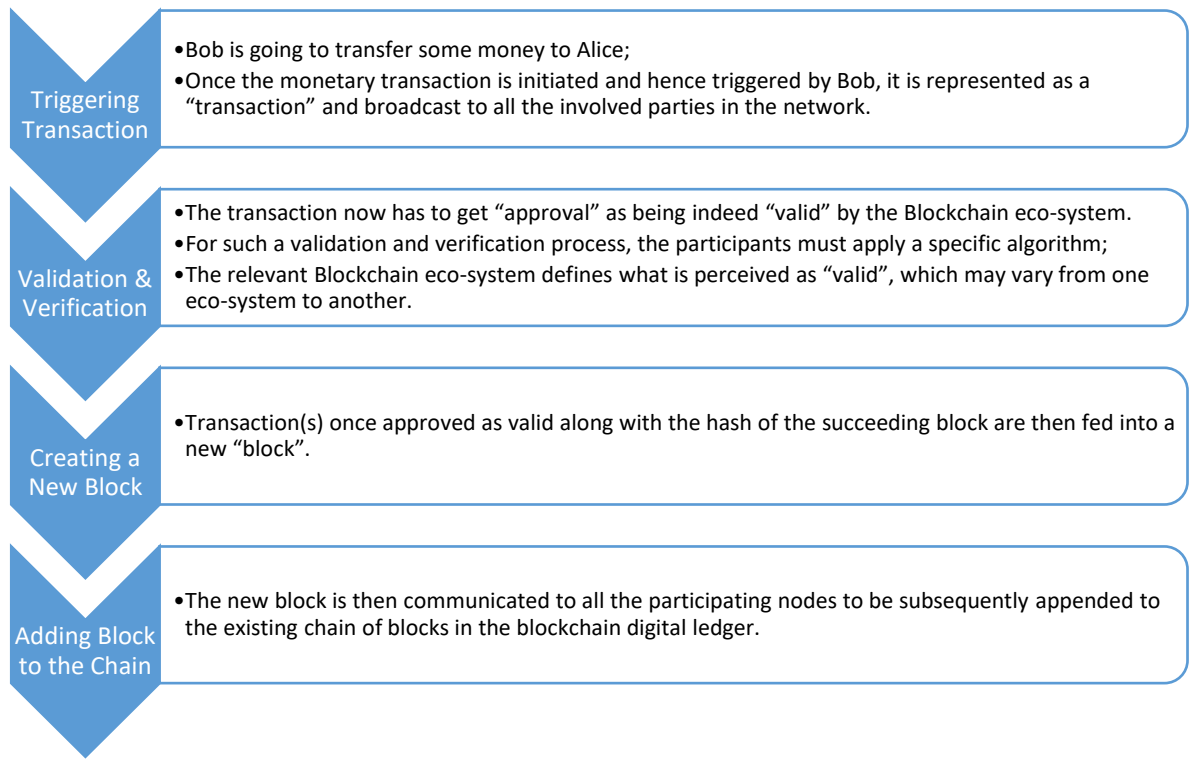

**Figure 1**. Operation of the Blockchain.

## 3. Use of Blockchain beyond Cryptocurrency

Although the Internet is a great tool to aid every sphere of the modern digital life, it is still highly flawed in terms of the lack of security and privacy, especially when it comes to FinTech and E-commerce. Blockchain, the technology behind cryto-currency, brought forth a new revolution by providing a mechanism for Peer-to-Peer (P2P) transactions without the need for any intermediary body such as the existing commercial banks [1]. BC validates all the transactions and preserves a permanent record of them while making sure that any identification related information of the users are kept incognito. Thus all the personal information of the users are sequestered while substantiating all the transactions. This is achieved by reconciling mass collaboration by cumulating all the transactions in a computer code based digital ledger. Thus, by applying Blockchain or similar crypto-currency techniques, the users neither need to trust each other nor do they need an intermediator; rather the trust is manifested within the decentralized network system itself. Blockchain thus appears to be the ideal "Trust Machine" [2] paradigm.

In fact, Bitcoin is just an exemplary use of the Blockchain. Blockchain is considered to be a novel revolution in the domain of computing enabling limitless applications such as storing and verifying legal documents including deeds and various certificates, healthcare data, IoT, Cloud and so forth. Tapscott [3] rightly indicated Blockchain to be the "World Wide Ledger", enabling many new applications beyond verifying transactions such as in: smart deeds, decentralized and/or autonomous organizations/ government services etc.

In the cloud [4,5] environment, the history of creation of any cloud data object and its subsequent operations performed thereupon are recorded by the data structure mechanism of 'Data Provenance', which is a type of cloud metadata. Thus this is very important to provide the utmost security to the data provenance for ensuring its data privacy, forensics and accountability. Liang *et al*. [6] puts forward a Blockchain based trusted cloud data provenance architecture, 'ProvChain', which is fully decentralized. Such adoption of the Blockchain in a cloud environment can provide strong protection against records being altered thus enabling an enhanced transparency as well as additional data accountability. This also increases the availability, trustworthiness, privacy and ultimately the value of the provenance data itself.

In an IoT ecosystem [7,8], most of the communication is in the form of Machine-to-Machine (M2M) interactions. Thus establishing trust among the participating machines is a big challenge that IoT technology still has not been met extensively. However, Blockchain may act as a catalyst in this regard by enabling enhanced scalability, security, reliability and privacy [9]. This can be achieved by deploying Blockchain technology to track billions of devices connected to the IoT eco-systems and used to enable and/or coordinate transaction processing. Applying Blockchain in the IoT ecosystem will also increase reliability by axing the Single Point of Failure (SPF). The cryptographic algorithms used for encryption of the block data as well as the hashing techniques may provide better security. However, this shall demand more processing power which IoT devices currently suffer from. Thus further research is required to overcome this current limitation.

Underwood [10] considers the application of Blockchain technology to completely overhaul the digital economy. Ensuring and maintaining trust is both the primary and initial concern of the application of the Blockchain. BC can also be used to gather chronological and sequence information of transactions, as it may be seen as an enormous networked time-stamping system. For example, NASDAQ is using its 'Linq Blockchain' to record its private securities transactions. Meanwhile the Depository Trust & Clearing Corporation (DTCC, USA) is working with Axoni in implementing financial settlement services such as post-trade matters and swaps. Regulators are also interested for BC's ability to offer secure, private, traceable real-time monitoring of transactions.

## 4. The Future of Blockchain

According to the Gartner Hype Cycle for Emerging Technologies 2017, shown in Figure 2, below, Blockchain still remains in the region of "Peak of Inflated Expectation" with forecast to reach plateau in "five to ten years". However, this technology is shown going downhill into the region of the "Trough of Disillusionment". Because of the wide adoption of the Blockchain in a wide range of applications beyond cryptocurrency, the authors of this paper are forecasting a shift in classification from "five to ten years" to "two to five years" to reach maturation. Blockchain possesses a great potential in empowering the citizens of the developing countries if widely adopted by e-governance applications for identity management, asset ownership transfer of precious commodities such as gold, silver and diamond, healthcare and other commercial uses as well as in financial inclusion. However, this will strongly depend on national political decisions.

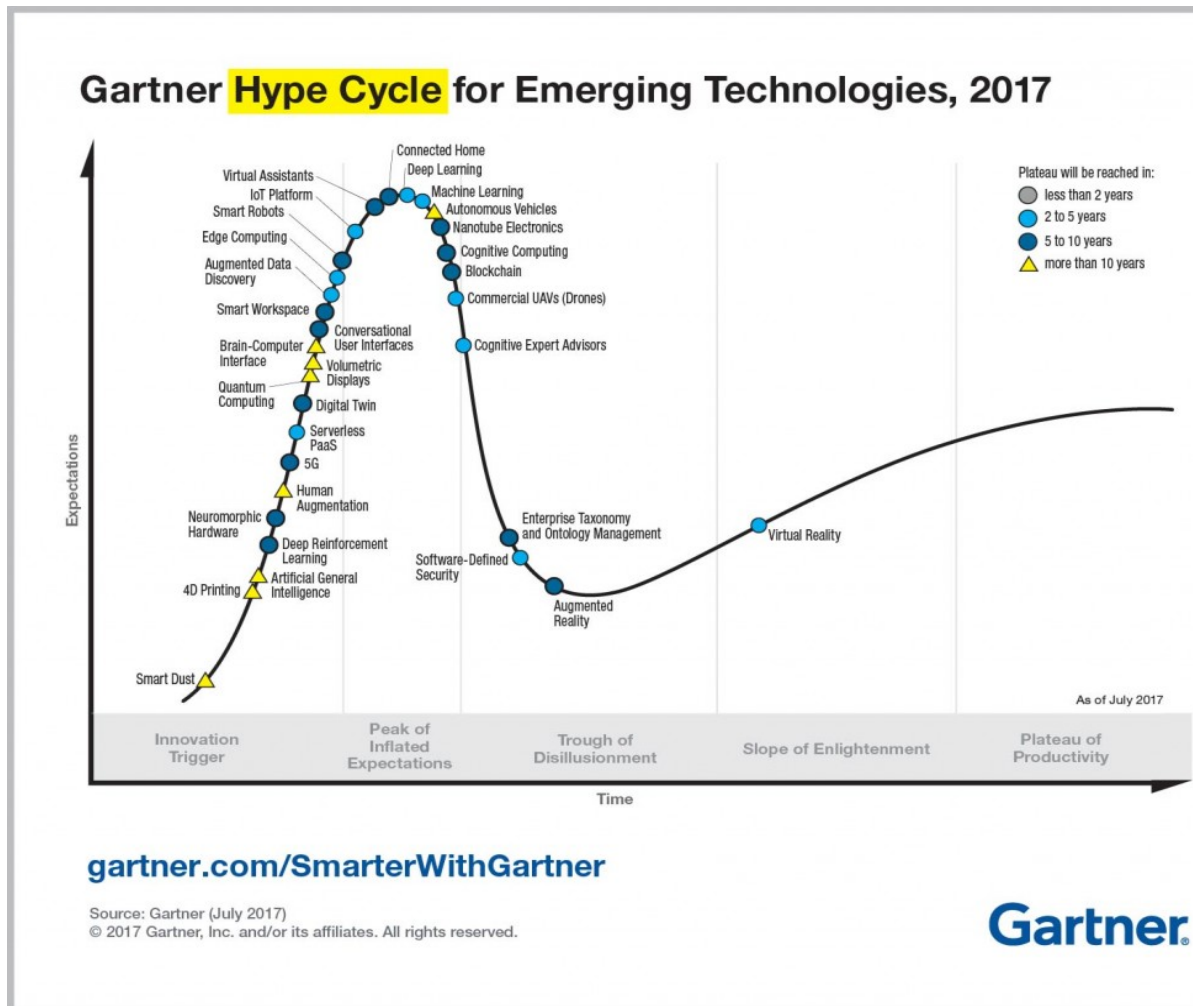

**Figure 2.** Gartner Hype Cycle, 2017 [11]

**6. Concluding Discussions**

The application of the Blockchain concept and technology has grown beyond its use for Bitcoin generation and transactions. The properties of its security, privacy, traceability, inherent data provenance and time-stamping has seen its adoption beyond its initial application areas. The Blockchain itself and its variants are now used to secure any type of transactions, whether it be human-to-human communications or machine-to-machine. Its adoption appears to be secure especially with the global emergence of the Internet-of-Things. Its decentralized application across the already established global Internet is also very appealing in terms of ensuring data redundancy and hence survivability.

The Blockchain has been especially identified to be suitable in developing nations where ensuring trust is of a major concern. Thus the invention of the Blockchain can be seen to be a vital and much needed additional component of the Internet that was lacking in security and trust before. BC technology still has not reached its maturity with a prediction of five years as novel applications continue to be implemented globally.

**References**

[1] Nir Kshetri, "Can Blockchain Strengthen the Internet of Things?," IT Professional, vol. 19, no. 4, pp. 68 - 72, May 2017, Available: http://ieeexplore.ieee.org/document/8012302/

[2] Mahdi H. Miraz, "Blockchain: Technology Fundamentals of the Trust Machine," Machine Lawyering, Chinese University of Hong Kong, 23rd December 2017, Available: http://dx.doi.org//10.13140/RG.2.2.22541.64480/2

[3] Don Tapscott and Alex Tapscott, Blockchain Revolution: How the Technology Behind Bitcoin Is Changing Money, Business, and the World, 1st ed. New York, USA: Penguin Publishing Group, 2016.

[4] Maaruf Ali and Mahdi H Miraz, "Cloud Computing Applications," in Proceedings of the International Conference on Cloud Computing and eGovernance - ICCCEG 2013, Internet City, Dubai, United Arab Emirates, 2013, pp. 1-8, Available: http://www.edlib.asdf.res.in/2013/iccceg/paper001.pdf

[5] Maaruf Ali and Mahdi H. Miraz, "Recent Advances in Cloud Computing Applications and Services," International Journal on Cloud Computing (IJCC), vol. 1, no. 1, pp. 1-12, February 2014, Available: http://asdfjournals.com/ijcc/ijcc-issues/ijcc-v1i1y2014/ijcc-001html-v1i1y2014/

[6] Xueping Liang et al., "ProvChain: A Blockchain-based Data Provenance Architecture in Cloud Environment with Enhanced Privacy and Availability," in Proceedings of the 17th IEEE/ACM International Symposium on Cluster, Cloud and Grid Computing (CCGrid '17), Madrid, Spain, May 14 - 17, 2017, pp. 468-477, Available: https://dl.acm.org/citation.cfm?id=3101176&CFID=994896989&CFTOKEN=44228545

[7] Mahdi H. Miraz, Maaruf Ali, Peter Excell, and Picking Rich, "A Review on Internet of Things (IoT), Internet of Everything (IoE) and Internet of Nano Things (IoNT)," in the Proceedings of the Fifth International IEEE Conference on Internet Technologies and Applications (ITA 15), Wrexham, UK, 2015, pp. 219 – 224, Available: http://ieeexplore.ieee.org/xpl/articleDetails.jsp?arnumber=7317398

[8] Mahdi H. Miraz, Maaruf Ali, Peter S. Excell, and Richard Picking, "Internet of Nano-things, Things and Everything: Future Growth Trends," (to be published) Future Internet, 2018.

[9] Mahdi H. Miraz and Maaruf Ali, "Blockchain Enabled Enhanced IoT Ecosystem Security," (accepted) in proceedings of the First International Conference on Emerging Technologies in Computing 2018 (iCETiC '18), London, UK, 23 August 2018.

[10] Sarah Underwood, "Blockchain Beyond Bitcoin," Communications of the ACM, vol. 59, no. 11, pp. 15-17, November 2016, Available: https://doi.org/10.1145/2994581

[11] Gartner, "Top Trends in the Gartner Hype Cycle for Emerging Technologies, 2017," Gartner, Inc., Gartner Hype Cycle 2017, August 2017, Available: http://www.gartner.com/smarterwithgartner/top-trends-in-the-gartner-hype-cycle-for-emerging-technologies-2017/